# Terrain Database Correlation Assessment Using an Open Source Tool

| | |
|---|---|
| **Leonardo Seiji Oyama, M.S.** | **Carlo Kleber da Silva Rodrigues, D.Sc.** |
| Brazilian Army | University Center of Brasília (UniCEUB) |
| Brasília, DF - Brazil | Brasília, DF - Brazil |
| oyama@coter.eb.mil.br | carlokleber@gmail.com |
| | |
| **Sérgio Simas Lopes Peres, M.S.** | **Brian Goldiez, Ph.D.** |
| Prospectare Brasil | University of Central Florida |
| Brasília, DF - Brazil | Orlando, FL |
| sp.ts.expert@outlook.com | bgoldiez@ist.ucf.edu |

## ABSTRACT

Configuring networked simulators for training military teams in a distributed environment requires the usage of a set of terrain databases to represent the same training area. The results of simulation exercises can be degraded if the terrain databases are poorly correlated. A number of methodologies for determining the correlation between terrain databases have been developed, by both government and industry, aiming at Verification, Validation & Accreditation of distributed simulations involving different simulators. However, there are few computational tools for this task and most of them were developed to address government needs, have limited availability, and handle specific digital formats. The goal of this paper is thus to present a novel open source tool developed as part of an academic research project. This tool analyzes a pair of terrain databases generating numeric data suitable for statistical analysis, as well as identifies specific areas where correlation may be an issue by using a configurable threshold. The analysis takes into consideration line-of-sight correlation differences between the databases. The sample size and characteristics of the line-of-sight tests, for instance elevation and azimuth, are selectable via a graphical user interface which also provides a 3D visualization of the terrain databases. Being open source, programmers may add more capabilities to the tool, such as including support to more digital formats or implementing new software methods to measure the correlation between terrain databases. Plans for extending the tool's capabilities and its possible utilizations are also included herein.

## ABOUT THE AUTHORS

**Leonardo Seiji Oyama**, M.Sc. holds the rank of Captain at the Brazilian Army. He has worked for 7 years at the Brazilian Army's Aviation providing technical support for helicopter flight simulators. His research interests include synthetic environments for distributed simulations and virtual reality. He holds a M.Sc. in Modeling & Simulation from the University of Central Florida and a B.S. in Computer Engineering from the Instituto Militar de Engenharia (Military Institute of Engineering) in Brazil. Mr. Oyama works as a simulation analyst at Simulation Division of the Land Operations Command.

**Carlo Kleber da Silva Rodrigues**, D.Sc. received the B.Sc. degree in Electrical Engineering (Federal University of Campina Grande in 1993), the M.Sc. degree in Systems and Computation (IME in 2000), and the D.Sc. degree in Systems and Computer Engineering (Federal University of Rio de Janeiro in 2006). He is professor at the University Center UniCEUB, in Brazil. Research interests include computer networks, multimedia applications, mathematical modeling, and cryptocurrencies.

**Sérgio Simas Lopes Peres**, M.Sc. has served in the Brazilian Army for over 32 years until he retired holding the rank of Colonel. He has worked at Simulation Division of the Land Operations Command for six years. He was the manager for the Armored Vehicle Guarani Simulation Project. Mr. Peres got his B.Sc. degree from the Military Academy of Agulhas Negras (Branch of Cavalry). He also has degrees in Business Management (Faculdades Integradas Amambai, 2003), Public Business Management (Universidade do Sul de Santa Catarina, 2010), Project Management (Fundação Getúlio Vargas, 2012), Training and Simulation Employment (Cranfield University, 2013), and Simulation for Acquisition (University of Central Florida, 2012). He holds a M.Sc. in Military Science (Advanced Army School). Currently he works at Prospectare Brasil as a simulation expert and consultant.

**Brian Goldiez**, Ph.D. is the Deputy Director at the University of Central Florida's Institute for Simulation and Training and a Research Associate Professor. Goldiez has over 40 years of experience in the modeling and simulation field spread over government, industry, and academia. His research focus has been optimizing the simulation system (human and simulation) through technological innovation. He has been most active over the past decade in simulation for healthcare and various aspects related to simulator interoperability, including LVC. Goldiez manages high performance computing resources at UCF and manages research related efforts in networking and visualization. Goldiez has degrees in Aerospace and Computer Engineering and a Ph.D. in Modeling and Simulation.

# Terrain Database Correlation Assessment Using an Open Source Tool

**Leonardo Seiji Oyama, M.S.**
Brazilian Army
Brasília, DF - Brazil
oyama@coter.eb.mil.br

**Carlo Kleber da Silva Rodrigues, D.Sc.**
University Center of Brasília (UniCEUB)
Brasília, DF - Brazil
carlokleber@gmail.com

**Sérgio Simas Lopes Peres, M.S.**
Prospectare Brasil
Brasília, DF - Brazil
sp.ts.expert@outlook.com

**Brian Goldiez, Ph.D.**
University of Central Florida
Orlando, FL
bgoldiez@ist.ucf.edu

## INTRODUCTION

There are numerous technical challenges to face when conducting a distributed simulation exercise involving distinct networked simulators (Goldiez, Salinas, Tarr, & Papelis, 2007). Beyond the issues regarding data exchange standards in use, the correlation between simulators' terrain databases (TDBs) deserves special attention (Joseph, Tosh, & Graniela, 2015). For instance, poorly correlated TDBs could cause unacceptable visual abnormalities, such as objects sinking into terrain or land vehicles floating above the ground.

When it comes to distributed military training exercises, interoperability issues caused by poor TDB correlation can be harmful because of adversely impacting on the outcomes of the engagement between simulated entities (Joseph et al., 2015). Line-Of-Sight (LOS) determination is one out of the many computations to calculate the simulation exercise outcomes which rely on the terrain and its features. When two entities are mutually visible in all networked simulators, there is an unblocked LOS between those entities. However, if there is an unblocked LOS between two entities in one of the networked simulators and those entities are not mutually visible in some other simulator in same the network, then there is a LOS correlation error between those entities. LOS calculations depend on the correlation between the TDBs. For example, an entity could get shot by an enemy that (from the entity's point of view) is totally covered by an elevation.

Within this context, this paper introduces the Runtime Terrain Database Correlation Assessment Tool (RTDBCAT). It is an open source tool that analyzes a pair of terrain databases, generating numeric data suitable for statistical analysis, as well as identifies specific areas where correlation may be an issue (Oyama, 2015). The analysis takes into consideration LOS correlation. Correlation test's parameters are selectable via a graphical user interface which provides a 3D visualization of the terrain databases. Being open source, there are no restrictions on its access and capabilities and enhancements may be added to the tool. As an example of potential enhancements is the support to a variety of digital formats and novel methods to measure correlation. With this in mind, plans for extending capabilities are thereby included at the end of this text.

One of the strengths of the methodology implemented in the tool is that it relies only on the geometry of the TDB, which must be composed of triangles. No assumptions about the purpose of the TDB were made, meaning that the tool can be used to address either visual or sensor TDBs. TDBs are used extensively in live, virtual, and constructive simulations and can include sensor data, logical data (for computer generated forces), and other terrain representations.

The remainder of this paper is structured as follows. In Fundamentals, a brief review of the main concepts related to military simulation training is given. Past Research includes a review of previous efforts in developing tools for assessment of correlation issues between TDBs. Then, it comes the main section of this paper: the novel assessment tool. It describes design and implementation details as well as provides insights of how the tool can be used to efficiently detect LOS correlation errors. The last section highlights the major limitations encountered during the elaboration of this work and unfolds the next steps aimed at eventual tool's improvement.

**FUNDAMENTALS**

Two guiding principles steered the research and prototype work described in this paper, namely interoperability and fair fight. Simulation interoperability is achieved when distinct simulators present the same behavior when given the same stimulation, considering a predetermined tolerance. To assess interoperability, it is necessary to apply the same stimulation to distinct systems and measure their corresponding behaviors, assuming a predefined number of statistic trials (Goldiez et al., 2007).

A fair fight is achieved when distinct simulators are interoperable and have similar performance capabilities considering a task over the entire simulation environment, within predefined tolerances. The definition of fair fight includes the similarity in the use of the synthetic environment, which encompasses the TDBs used in the simulation (Goldiez et al., 2007). Therefore, a set of correlated TDBs is a necessary, but not sufficient condition for a fair fight in distributed simulations and any tool created to assess correlation should be open source, so that it can be properly vetted and enhanced by others.

Still, the terms "virtual" and "constructive" are concepts related to the military taxonomy in training simulations According to the Modeling & Simulation Coordination Office Glossary (2016), the former is a simulation that involves real people using simulated equipment, and the latter is a simulation that involves simulated people using simulated equipment. Since the RTDBCAT relies uniquely on the TDB's geometry, it can address TDBs created for both virtual and constructive simulations as long as they are composed of triangles.

**PAST RESEARCH**

Interoperability issues caused by poor TDB correlation has become a constant challenge when conducting distributed simulation exercises because a multitude of simulators and IG software were not designed for interoperability (Schiavone & Goldiez, 2000). In spite of the numerous existing literature works, TDB correlation analysis is still considered as an issue that deeply concerns private industry and, more especially, government organizations (Coad, Crush, Page, & Smith, 2016). Within this context, in what follows we comment on several main researches and tools addressing this topic since the 90's.

Line of Sight Intervisibility Metrics (LOSIM) (Hoffman, Horan, McDonald, Paris, & Uliano 1994) was research conducted in the early 90's focused on the differences of the rendered images of a database. The authors developed a metric to assess correlation between two TDBs based on intervisibility. Through that metric, it was possible to calculate the extent to which two different image generators would provide the same LOS intervisibility for TDBs representing the same environment. LOSIM relies on statistic tests to determine how correlated are two TDBs regarding LOS intervisibility. It uses test pairs instead of test points for reducing the sample size without losing statistic power. LOSIM also utilizes human factors to reduce sampling size (for example, naked human eye viewing threshold). LOSIM is the oldest tool encountered in this review of the literature, followed by ZCAP.

ZCAP (Sakude, Schiavone, Morelos-Borja, Martin, & Cortes 1998; Schiavone, Sakude, Graniela, Morelos-Borja, & Cortes, 1998) is a set of tools for TDB correlation analysis capable of performing a number of correlation tests, namely elevation, LOS, culture, shift detection, and visualization. ZCAP was developed at the University of Central Florida Institute for Simulation and Training and funded by the former United States (US) Army Simulation, Training and Instrumentation Command (now US Army PEO-STRI) and the former Defense Modeling and Simulation Office (DMSO). ZCAP stands for Z-Correlation Analysis Program. It was originally designed for assessing elevation correlation between TDBs, but, more functionalities were added to it later, including the ability to handle several TDB formats used by the United States Army. As of October 1998, ZCAP was capable of performing the following tests: Elevation Correlation, Line-of-Sight Correlation, Culture Correlation, Visual Correlation and Shift Detection (Schiavone et al., 1998). Besides ZCAP, another important TDB correlation research conducted by the US Government was SEE-IT.

SEE-IT (Sedris.org, 2014) is a tool developed by SEDRIS. SEDRIS was sponsored by the DMSO and other US government organizations. SEDRIS did a noteworthy effort in the sense of creating a neutral TDB format. SEE-IT is capable of performing consistency tests within TDBs and correlation analysis between TDBs, but it can handle SEDRIS native TDB format only. The utilization of SEE-IT requires one or more TDB conversion processes. SEDRIS also created the Spatial Reference Model for converting system coordinates, which was reused by the LightBox years later.

Another TDB correlation tool created in the early 2010's is the LightBox (Palmer & Boyd, 2011). It is a tool capable of performing automated correlation analysis that relies on the utilization of Graphical Processing Units (GPU) to accelerate tests. This tool can accept a number of TDB formats and can perform a multitude of correlations tests. The idea of using GPU to accelerate correlation tests was previously presented by Tracy (2004). Another remarkable feature of LightBox is its sampling method. The sample points for detecting LOS correlation issues are placed close to terrain features, such as buildings and trees. The LightBox development was sponsored by SAIC. It is the penultimate tool encountered in archival sources, being "Fit-for-use" the last one.

"Fit-for-Use" (Joseph, Tosh, & Graniela, 2015) is a tool developed to address the needs of the US Navy, Naval Air Systems Command (NAVAIR). When it was created, none of the existing tools could support NAVAIR Portable Source Initiative (NPSI). This tool is able to perform tests between sensor TDBs and visual TDBs, and also within visual TDBs, including several formats in use in the NAVAIR. According to Joseph et al. (2015), the tool can perform several correlation and integrity tests on elevation, imagery and terrain culture and can handle TDBs containing large amounts of geo-specific imagery.

Finally, the last tool mentioned in this section is C-nergy. C-nergy is a framework for TDB correlation and integrity testing developed by the Dignitas Technologies LLC (Dignitas Technologies – M&S Tools, 2017). According to advertisements, C-nergy natively supports OpenFlight format and OneSAF Objective Terrain Format, but the announcement also states that more formats and correlation tests can be added by developers via plugins. Many native correlation and integrity tests in C-nergy are focused on OneSAF, which is the US Army standard simulator for computer generated forces.

The list of tools presented herein is a partial summary and non-exhaustive. Only work found on archival sources and Internet search were included in this paper. Considering this brief review, one can note that previous work addressed specific needs of government organizations. Even those tools created by private companies are not accessible or cannot be distributed due to contractual clearance guidelines. On the other hand, the tool described herein addresses this gap. It is not focused on specific needs of any organization and because its source code is available, others can add functionalities to it according to their own needs. One of the strengths of the current project over the other efforts is that it is entirely is published under public domain, allowing other researchers to freely modify not only the core functions of the software, but also its interface. The technique also builds upon the work of others, but provides a unique sampling method that is statistically sound. The work being open source and public allows others to enhance it over time.

**THE NOVEL CORRELATION ASSESSMENT TOOL**

This section introduces the RTDB correlation assessment tool and is structured as follows. The first subsection discusses the tool's technology and architecture. The second subsection describes how the user should start using the tool and the expected results from the tool's initialization. The third subsection describes how the tool execute the LOS tests. The fourth subsection clarifies how the tool calculates the roughness of a TDB. The fifth subsection describes how the test data can be processed, analyzed, and visualized.

**Technical Features and Architecture**

The RTDBCAT was written in the C++ object-oriented programming language along with the Qt framework. The C++ language has been selected for two reasons. The first reason is that many application programming interfaces (APIs) used to handle TDBs are written in C++. The second reason is that C++ allows the control of computer memory, which is useful when handling large databases. The Qt framework has been chosen because it can be smoothly integrated with most C++ APIs, it is opens source, and it facilitates the creation of graphical user interfaces (GUIs). The tool is composed of three modules. At the present time, all the modules are statically linked.

The first module is responsible for accessing the TDBs and for converting coordinates to/from different reference systems. This module is also responsible for preparing the data that is displayed on the second module. The second module is basically the tool's GUI. For example, the RTDBCAT visualization windows are part of the second module. The visualization windows, which are used to display the TDBs, are based on the QGLWidget class. All the other elements comprised in the GUI are based on the QWidget class.

Lastly, the third module is responsible for the tool's heavy computational work. The line-to-triangle intersection determination and terrain roughness computation algorithms, which will be discussed later in this paper, are part of this module. These algorithms are Central Processing Unit (CPU) centric. The third module is also responsible for recording the numeric data from the tests in the computer hard disk.

**Initialization**

The tool was originally created to demonstrate a methodology on assessment of TDB correlation using LOS measurements (Oyama, 2015), therefore, its GUI is intended to facilitate the insertion of parameters used to perform LOS tests. As mentioned before, the RTDBCAT compares a pair of TDBs. Once they are selected, the tool displays textual information about the terrain geometry, such as number of vertices, polygons, and levels of detail (LODs) present in each TDB, as well as geographic information, such as latitude, longitude, and elevation extents.

Before proceeding to the correlation test, the user must select which LOD should be used in the assessment. For instance, the user may want to select the most detailed representation of each TDB to perform the correlation tests. Once this decision is made, the tool will enable visualization windows, allowing the user to see both TDBs. These visualization windows are not used to generate statistical data. Analysis based on pure human visual inspection may be inaccurate, however visual feedback can be useful to inspect correlation issues detected in automated tests.

By using the visualization windows, the user can move and rotate the virtual camera freely through the TDBs, so that it is possible to visualize any portion of the databases. Figure 1 shows corresponding portions of two different TDBs representing the same area of interest (AOI) depicted from the visualization windows. Figure 1 (a) shows a TDB created from a digital elevation model (DEM) with 1 arc-second resolution, whilst the Figure 1 (b) shows a TDB created from a DEM with 1/3 arc-second. Both databases represent the same location in the State of California. The position and orientation of the virtual camera is exactly same for both images. Considering these images, it is possible to note that there are differences between the TDBs, but it is not possible to assess how strong is the correlation between them without performing the automated tests.

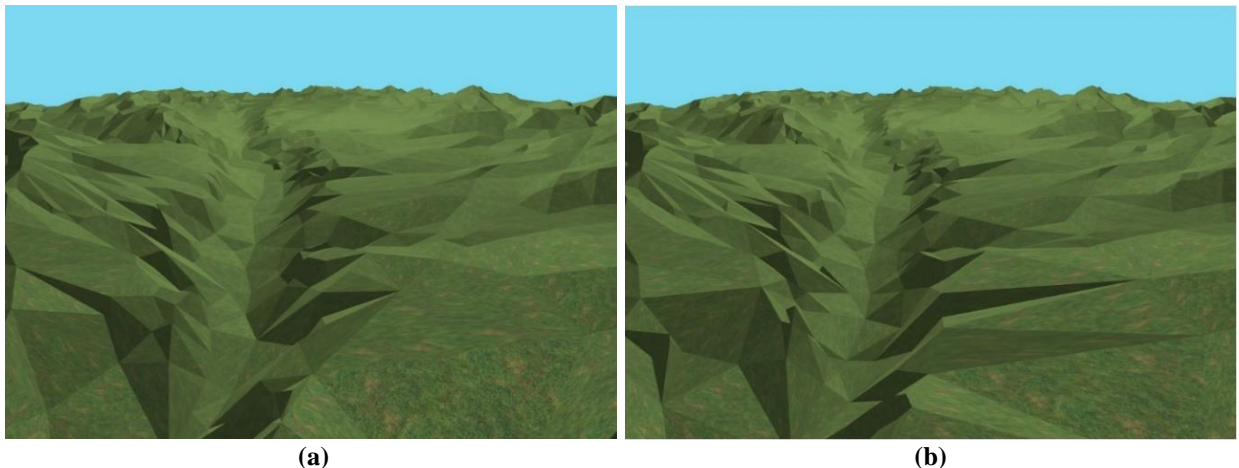

(a)                                                                                          (b)
**Figure 1. (a) TDB created from 1arc-second DEM, (b) TDB created from 1/3 arc-second DEM**

The RTDBCAT allows the user to divide the TDBs into smaller portions of nearly equal area called terrain blocks. The advantage of dividing the TDBs is that this allows the detection of poorly correlated parts of terrain. Before starting the tests, the user must indicate how the TDBs will be divided by specifying a quantity of lines and columns. The user can set both values to 1 in order to avoid dividing the TDB.

**LOS Correlations Tests**

The assessment of LOS correlation implemented in the RTDBCAT relies on the differences between the lengths of corresponding line segments in both TDBs. These line segments are called LOS rays. The starting point of a LOS ray is called eyepoint. The direction of any line segment, such as a LOS ray, can be described by one directional unit vector (vector of length 1).

In order to ensure that all the LOS rays have a measurable length, the tool encapsulates the TDBs inside a delimiting volume called bounding box. The LOS rays that miss the terrain surface will intersect the "fictional invisible wall" or the "fictional invisible ceiling" of the bounding box. The absence of bounding boxes would break the tool's correlation assessment method, because the tool utilizes the lengths of the rays to compute correlation.

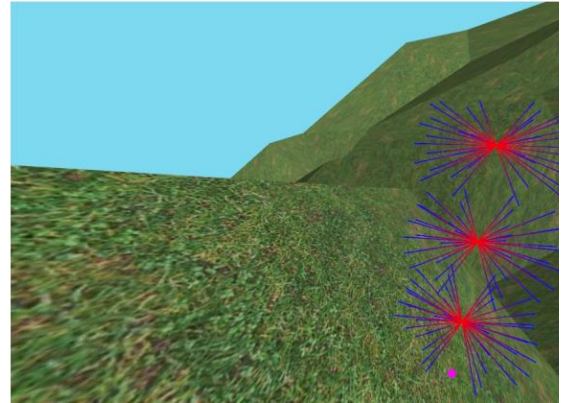

**Figure 2. Example of LOS Test Configuration**

After dividing the TDBs into smaller portions, the user must determine how many locations per terrain block will be tested. The test locations are distributed evenly within each terrain block. One or more eyepoints can be placed on each test location at some level above the terrain surface, according to the user specification. The user also must specify how many eyepoints will be placed at each test location, the vertical distance between the eyepoints, and the distance above ground level (AGL) of the eyepoint which is closest to the terrain surface. After this, the user must define the LOS rays that will be traced from each eyepoint. To accomplish this, the user must specify the directional vectors that will define the direction of the LOS rays, by specifying the horizontal and vertical angular variations (azimuth and pitch angles) of the directional vectors. Figure 2 illustrates the configuration corresponding to 3 eyepoints per test location along with 24 directional vectors per eyepoint. The small magenta point on the ground is a sample test location and it also represents the vertical projection of the 3 eyepoints.

The computation to determine if a LOS ray intersects the terrain surface involves parametric equations of lines and planes, and a line-to-triangle intersection algorithm based on linear algebra. All the triangles composing the terrain surface must be tested, however rear facing triangles with respect to the LOS ray are excluded from the intersection test. Oyama (2015) describes the math behind these algorithms in detail. When the tool finishes the LOS tests, the resulting LOS rays can be shown in the visualization windows. Figure 3 shows two corresponding LOS rays traced in their respective TDBs. The LOS depicted in the Figure 3 (a) intersects the terrain surface, whilst its counterpart depicted in the Figure 3 (b) is intersecting the bounding box's "invisible wall".

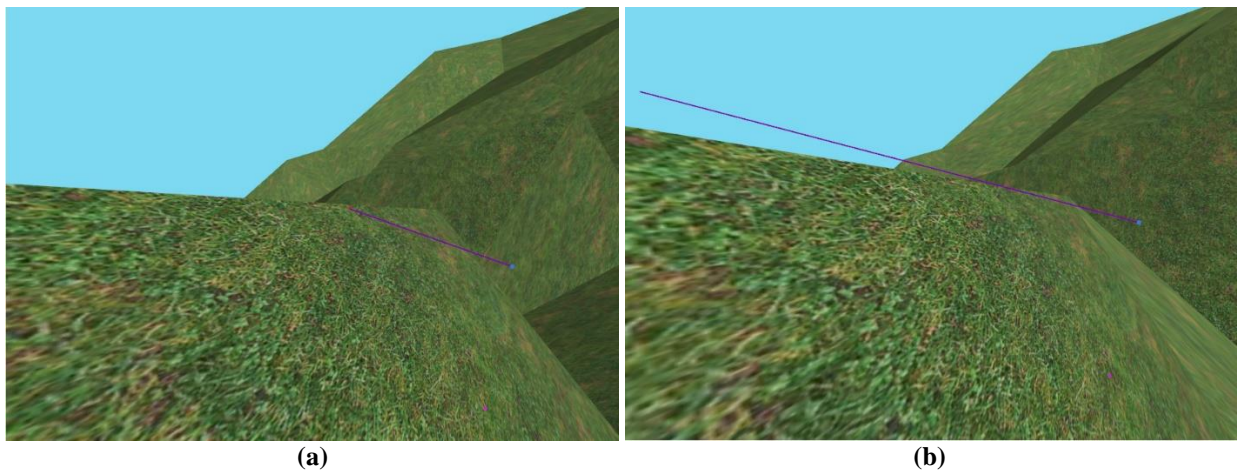

(a)          (b)
**Figure 3. (a) LOS ray intersecting the terrain surface, (b) LOS ray intersecting the bounding box**

**Calculation of Terrain Roughness**

Besides performing LOS tests, the tool is also capable of measuring the roughness of TDBs and its subdivisions. It's reasonable to believe that rough parts of the TDBs are more likely to present LOS correlation errors than flat parts. Thus, rough areas should be subject to further investigation. Additionally, differences between the roughness of the selected TDBs indicate possible LOS correlation errors, therefore impacting interoperability and fair-fight issues. For instance, if only one of the selected TDBs is highly rough, it is clear that this pair of TDBs is not amenable to the task of shooting direct fire weapons.

The method for determining terrain roughness proposed by Oyama (2015) is intended to calculate the roughness of any TDB composed uniquely of triangles. This method utilizes the unit vectors normal to the triangles that constitute the TDB. The general idea of this technique is measuring the "intensity" of the 3D dispersion of the vectors normal to the triangles that integrate the TDB.

Oyama (2015) proposes that the roughness for a TDB composed of *N* triangles, where the vector ($x_i$, $y_i$, $z_i$) represents the vector normal to the *i-th* triangle in the TDB, is achieved through Equations (1), (2) and (3). Equation (1) is similar to the equation proposed by Hobson (1972) to calculate the *vector strength*, but it is slightly different, except for the denominator, *N*. Equation (1) is equivalent to the "average" of the vectors normal to the TDB's triangles.

$$(\bar{x}, \bar{y}, \bar{z}) = \left( \frac{\sum_{i=1}^{N} x_i}{N}, \frac{\sum_{i=1}^{N} y_i}{N}, \frac{\sum_{i=1}^{N} z_i}{N} \right) \quad (1)$$

Equation (2) calculates the 3D dispersion of vectors normal to the triangles that are part of the TDB. This equation is equivalent to the "standard deviation" of the vectors normal to the TDB's triangles.

$$(\sigma_x, \sigma_y, \sigma_z) = \left( \sqrt{\frac{\sum_{i=1}^{N}(x_i - \bar{x})^2}{N}}, \sqrt{\frac{\sum_{i=1}^{N}(y_i - \bar{y})^2}{N}}, \sqrt{\frac{\sum_{i=1}^{N}(z_i - \bar{z})^2}{N}} \right) \quad (2)$$

Finally, the Equation (3) provides the TDB's roughness value, which is precisely the length of the vector obtained by the Equation (2).

$$|(\sigma_x, \sigma_y, \sigma_z)| = \sqrt{\sigma_x^2 + \sigma_y^2 + \sigma_z^2} \quad (3)$$

The roughness value varies from 0 to 1, where 0 means completely flat. One of the strengths of this roughness calculation method is that it is not affected by the terrain's overall steepness.

**Data Analysis and Visualization**

The user can save the results of LOS tests and TDB roughness calculation into a file in the computer's hard disk. This file contains human readable numeric data that can be used for later statistical analysis. The numbers are separated by a tab character ('\t' character in C programming language). Currently, the tool is not integrated to any statistical software, therefore the resulting file is the only way for accessing the numeric data produced at the end of the tests.

Statistical software can provide help for generating histograms and Q-Q plots, as well as they can help the user to perform normality tests on the data contained in the RTDBCAT output file, such as Kolmogorov-Smirnov, Anderson-Darling, and Shapiro-Wilk (Dragulescu, 2014; Meyer, Dimitriadou, Hornik, Weingessel, and Leisch, 2015; Gross & Ligges, 2015; Lemon, 2006; R Core Team, 2015; Verzani, 2015). Other statistic measures, such as minimum, maximum, average, trimmed average, median, standard deviation, standard error, skewness, and kurtosis, can be easily calculated with the help of specialized statistical software.

In addition to the RTDBCAT, an auxiliary tool was created as part of the same academic research. This prototype tool can generate Microsoft Excel files (XLSX file extension) that ease the visualization of the tests recorded in the RTDBCAT output file. This tool can color cells in Excel sheets according to thresholds specified by the user. Figure 4 illustrates a block to block roughness comparison between TDBs selected by the user. In this sample, the auxiliary tool highlighted in green terrain blocks with roughness value lesser than 0.25, whilst it highlighted in red terrain blocks with roughness value greater than 0.50. If there was a terrain block highlighted in green in one of the TDBs and highlighted in red in the other TDB, it could be interpreted as a possible problem for interoperability. No attempt was made to delineate the highlighting threshold for the auxiliary prototype tool. A 'good enough' value is situationally dependent. It is hoped that over time a set of thresholds will be created based on the needs of different user communities.

|      | North |  |  |  |  |      |
|------|-------|-------|-------|-------|-------|------|
|      | Block 4_0 | Block 4_1 | Block 4_2 | Block 4_3 | Block 4_4 |      |
|      | 0.526841 | 0.577333 | 0.587508 | 0.527934 | 0.205784 |      |
|      | 0.537245 | 0.58108 | 0.600487 | 0.528962 | 0.222668 |      |
|      | Block 3_0 | Block 3_1 | Block 3_2 | Block 3_3 | Block 3_4 |      |
|      | 0.55615 | 0.550422 | 0.489808 | 0.619079 | 0.412622 |      |
|      | 0.597579 | 0.567785 | 0.521004 | 0.635282 | 0.413921 |      |
| West | Block 2_0 | Block 2_1 | Block 2_2 | Block 2_3 | Block 2_4 | East |
|      | 0.543875 | 0.54719 | 0.433546 | 0.619856 | 0.557562 |      |
|      | 0.553922 | 0.556558 | 0.455554 | 0.637175 | 0.563058 |      |
|      | Block 1_0 | Block 1_1 | Block 1_2 | Block 1_3 | Block 1_4 |      |
|      | 0.557646 | 0.502785 | 0.524551 | 0.588068 | 0.600436 |      |
|      | 0.576867 | 0.504986 | 0.533105 | 0.601528 | 0.609928 |      |
|      | Block 0_0 | Block 0_1 | Block 0_2 | Block 0_3 | Block 0_4 |      |
|      | 0.494554 | 0.408531 | 0.521889 | 0.369569 | 0.429209 |      |
|      | 0.504008 | 0.429738 | 0.534979 | 0.380793 | 0.447182 |      |
|      | South |  |  |  |  |      |
|      | Highlighted in green: 0.00 < (Roughness) < 0.25 |  |  |  |  |      |
|      | Highlighted in red: 0.50 < (Roughness) < 1.00 |  |  |  |  |      |

**Figure 4. Block Roughness Comparison**

**Tool's Usefulness**

This tool has utility from several points of view. First, it can be used to help in scenario generation by indicating areas where correlation is high and therefore where interactions are less likely to be impacted by interoperability issues. Conversely, if there are specific areas where correlation is important to training and where problems in correlation exist, the tool can point to areas where analysts need to spend time addressing the correlation problem by adjusting detail and polygon distribution. Secondly, an open tool invites others to freely use the tool, identify problem areas or shortcomings, and contribute to making the tool better for the general benefit of the modeling and simulation community. For example, commercial airline simulators can benefit from a tool such as this when correlating airfield data bases with actual terrain. Similarly, the design of roadways and airfields can benefit when using a tool such as has been described in this paper to look at sight lines for visibility and aesthetics.

**LIMITATIONS AND FUTURE WORK**

**Architecture Limitations**

RTDBCAT needs a major architecture change in order to replace the existing static linkage between its modules. Dynamically linked modules could facilitate the replacement and the addition of functionalities, such as the input of data from the hard disk. The development of a plugin system could help to improve the tool's modularity.

At the present time, the RTDBCAT does not include a statistical module. Existing statistical open source software libraries could extend the tool's functionality, so that statistical analysis could be performed without the need of external tools. To achieve this, the RTDBCAT's architecture should evolve to include a new module for statistics.

RTDBCAT can only assess correlation between two triangulated TDBs. This limitation could be surpassed by including triangulation algorithms in the tool. However, some IG software can render images directly from elevation posts, which means that the terrain is triangulated at runtime. In such case, there would be no way to guarantee the quality of the correlation assessment, because it could not be possible to ensure that the tool's triangulation algorithm matches the IG triangulation algorithm.

**Support to Additional TDB formats**

In 2018, the Brazilian Army (BA) will take part in the Viking computer assisted exercise, conducted by the Swedish Armed Forces. Many countries will participate on Viking 2018, which makes room to the evolution and usage of the RTDBCAT as a support to decision instrument. However, the tool needs improvements to be used in such simulation environment, such as the inclusion of support to more TDB formats.

Currently, the tool is capable of handling only one TDB format. This format has been chosen because it is used in two simulation systems at the BA, namely the Aviation Flight Simulation and the Artillery's Fire Support Simulation. The inclusion of support to more TDB formats would require modifications in the module of RTDBCAT responsible for retrieving information from the TDBs. Future tool improvements possibly should include support to TDB formats used in BA simulators and serious games, such as SWORD, VBS3, Steel Beasts, and more. However, the inclusion of support to additional TDB formats is a challenging task, because of the absence of friendly APIs to handle some TDBs. Often TDB developers provide only the binary file specification of their databases, which can change in future versions of the simulator without previous notification.

Because the code and documentation of the tool are open source and public others can to adapt it to different terrain format (and hopefully return the enhancements to serve the public).

**Additional Correlation Tests**

Besides the correlation tests currently supported, more tests must be added to the RTDBCAT to expand its usability. New correlation tests include vector correlation analysis. The implementation of vector correlation test would allow the detection of misplaced terrain features. Vector correlation analysis is a very useful test, because the misplacement of rivers, lakes, railways, highways and roads is a common issue. Vector correlation analysis should also check for a vector's attributes with respect to correlation. For example, a vector that represents a road in one of the TDBs could be wrongly representing a highway in another TDB.

To date, the RTDBCAT is not capable of comparing 3D models. The implementation of correlation analysis of 3D models should include analysis of a multitude of parameters, such as position, orientation, polygon count, LOD count, and texture.

**Performance Improvements**

The algorithms implemented in the RTDBCAT are CPU centric. Existing techniques make use of GPU parallelism to accelerate intense and repetitive computations. To date, GPU centric computing techniques can be implemented in graphics card independent programming language, making the tool available to larger number of users. The implementation of GPU centric algorithms will require modifications to the RTDBCAT's second module.

**CONCLUSION**

The goal of this paper was to describe the RTDBCAT, which is the first known open source tool for TDB correlation assessment. This paper discussed the RTDBCAT's architecture, its utilization workflow, and its currently implemented correlations tests (LOS and terrain roughness). The tool's limitations were described, as well as eventual development plans improve its capabilities. The tool is released under public domain and, thereby, organizations and individuals can make use of it "as it is" or implement modifications to the tool's source code to meet specific requirements.

**Disclaimer**

The views expressed in this paper are those of the authors and do not necessarily represent the official policy or position of the organizations with which they are affiliated.